\documentclass[runningheads]{llncs}
\usepackage[T1]{fontenc}
\usepackage{graphicx}
\usepackage[colorlinks=true,linkcolor=blue,citecolor=blue,urlcolor=blue]{hyperref}
\usepackage{url}

\usepackage{tikz}
\usepackage{float}
\usepackage{amssymb}
\usepackage{amsmath}
\usepackage{pgfplots}
\pgfplotsset{compat=1.18}
\usetikzlibrary{positioning,decorations.pathreplacing,fit,backgrounds,patterns}

\begin{document}
\title{Layered Performance Analysis of TLS 1.3 Handshakes: Classical, Hybrid, and Pure Post-Quantum Key Exchange}

\titlerunning{Performance Analysis of TLS 1.3 PQ Handshakes}
\author{David G\'omez-Cambronero\inst{1}\orcidID{0009-0006-4127-1853} 
Daniel Munteanu\inst{2}\orcidID{0009-0005-1982-3706} 
Ana I. Gonz\'alez-Tablas \inst{3}\orcidID{0000-0002-6259-8955}
}
\authorrunning{D. G\'omez-Cambronero et al.}
%
\institute{Telefonica Innovaci\'on Digital, Madrid, Spain \\
\email{david.gomezcambronero@telefonica.com}
\and
Keysight Technologies, Bucharest, Romania \\
\email{daniel.munteanu@keysight.com}
\and
Universidad Carlos III de Madrid, Spain \\
\email{aigonzal@inf.uc3m.es}}
\maketitle              
\begin{abstract}
In this paper, we present a laboratory study focused on the impact of post-quantum cryptography (PQC) algorithms on multiple layers of stateful HTTP over TLS transactions: the TCP handshake, the intermediate TCP–TLS layer, the TLS handshake, the intermediate TLS layer, and the HTTP application layer. To this end, we propose a laboratory architecture that emulates a real-world setup in which a load test of up to 100 transactions per second is sent to a load balancer, which in turn forwards them to a backend server that returns the responses.
Each set of tests is executed using the TLS 1.3 key exchange groups as follows: traditional (or non-PQC), hybrid PQC and pure PQC. Each set of tests also varied the backend response size.
Across more than thirty experiments, we performed data reduction and statistical analysis for each layer, to determine the specific impact of each algorithm (PQC and traditional) at every stage of the HTTP-over-TLS transaction.

\end{abstract}
\section{Introduction}

The advent of cryptographically relevant quantum computers challenges the security assumptions of current public-key infrastructures. In response, NIST finalized its first post-quantum cryptographic (PQC) standards in August 2024: FIPS 203 (ML-KEM)~\cite{nist_mlkem}, FIPS 204 (ML-DSA)~\cite{nist_mldsa}, and FIPS 205 (SLH-DSA)~\cite{nist_slhdsa}. Growing concern over “store now, decrypt later” attacks has accelerated PQC migration efforts, making it necessary to quantify performance impact prior to deployment.

While previous studies have evaluated PQC at the aggregate TLS handshake level, this work introduces a layered decomposition of the HTTP-over-TLS transaction into five protocol phases (3 typical TLS layers: TCP, TLS, App, and the intermediate TCP-TLS and TLS-App, see Figure~\ref{fig:tls13_handshake}), enabling attribution of PQC-induced overhead to specific layers.

The main contributions of this paper are:
\begin{enumerate}
  \item A five-layer latency decomposition methodology for TLS~1.3 connections, from TCP handshake through application response.
  \item An empirical comparison of classical key exchange (x25519), hybrid (x25519+ML-KEM), and pure post-quantum (ML-KEM) under realistic load conditions (5 minutes with 100~TPS) in more than 30~performance tests (in total 30K requests per each performance test with a total of around 1 million requests).
  \item The finding that the TLS handshake \emph{exchange} (ClientHello$\to$Finished) is effectively \emph{algorithm-neutral}: all configurations---classical, hybrid, and pure ML-KEM---show negligible effect sizes (Glass's $\Delta<0.2$--$0.33$), with no practically meaningful penalty or advantage attributable to the key exchange algorithm. The algorithm-sensitive cost is instead isolated to ClientHello \emph{construction}, which we measure separately in the TCP-to-TLS layer; this finding is consistent with Sosnowski et al.~\cite{acm2024}.
  \item A publicly available data-reduction tool for extracting per-layer statistics from packet captures.
\end{enumerate}

The remainder of this paper is organized as follows. Section~2 reviews background and related work. Section~3 describes the methodology and experimental setup. Section~4 presents and discusses the results. Section~5 concludes and outlines future work.

\section{Background and Related Work}

Throughout this paper, the term \emph{classical} (or \emph{traditional}) refers to cryptographic algorithms that are vulnerable to attacks by quantum computers (e.g., ECDH, RSA), while \emph{post-quantum} (PQC) refers to algorithms specifically designed to be resistant to such attacks (e.g., ML-KEM). Note that all algorithms---classical and post-quantum alike---run on conventional (non-quantum) hardware; the distinction is solely about their vulnerability to quantum adversaries~\cite{bernstein2009}.

\subsection{PQC Standardization and Migration Context}

Following the NIST PQC standards finalized in August 2024~\cite{nist_mlkem,nist_mldsa,nist_slhdsa}, and their adoption by software providers such as the Open Quantum Safe (OQS) project~\cite{oqs2024} and OpenSSL~\cite{openssl2024}, network vendors are starting to integrate these algorithms into their products. Furthermore, considering the threat of ``store now, decrypt later'' attacks, organizations are paying close attention to quantum-safe cryptography and building PQC transition initiatives. 

The urgency of this transition is underscored by institutional mandates. The U.S.\ National Security Agency published updated guidance for the CNSA~2.0 algorithm suite~\cite{cnsa2}, requiring all national security systems to adopt ML-KEM for key establishment by 2030 and quantum-resistant algorithms exclusively by 2033. These timelines, together with similar initiatives in the European Union and other jurisdictions, create a pressing need for empirical performance data to inform migration planning. Therefore, for all stakeholders, it is of paramount importance to understand the performance impact on their systems when migrating to PQC before pushing this into production.

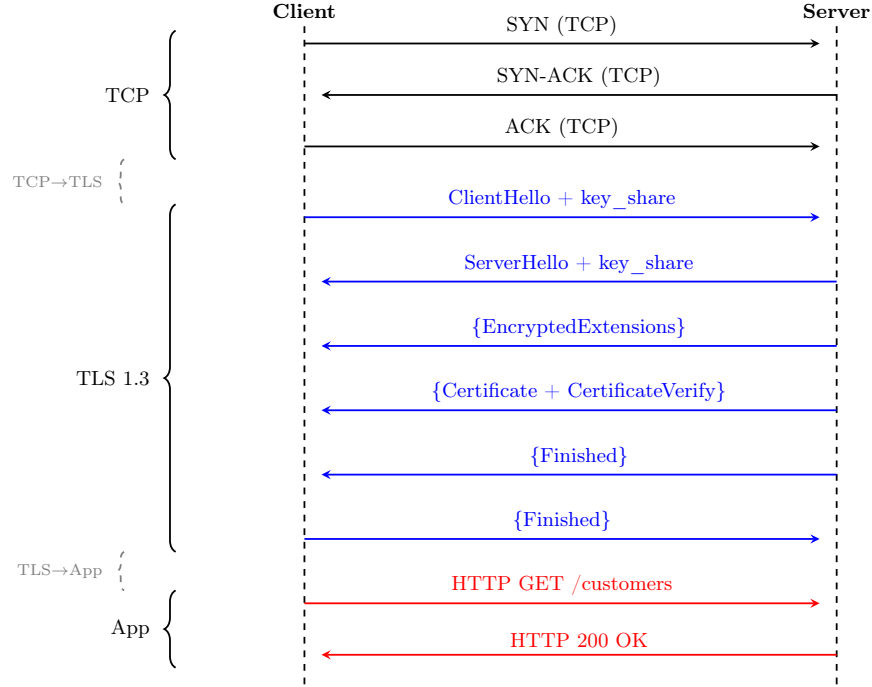
\begin{figure}[H]
\centering
\resizebox{0.95\textwidth}{!}{%
\begin{tikzpicture}[>=stealth, thick]
  \node (client) {\textbf{Client}};
  \node[right=7cm of client] (server) {\textbf{Server}};
  \draw[dashed] (client) -- ++(0,-10.5);
  \draw[dashed] (server) -- ++(0,-10.5);
  \draw[->] (client)++(0,-0.5) -- ++(8,0) node[midway, above]{\small SYN (TCP)};
  \draw[->] (server)++(0,-1.3) -- ++(-8,0) node[midway, above]{\small SYN-ACK (TCP)};
  \draw[->] (client)++(0,-2.1) -- ++(8,0) node[midway, above]{\small ACK (TCP)};
  \draw[->, blue] (client)++(0,-3.2) -- ++(8,0) node[midway, above]{\small ClientHello + key\_share};
  \draw[->, blue] (server)++(0,-4.2) -- ++(-8,0) node[midway, above]{\small ServerHello + key\_share};
  \draw[->, blue] (server)++(0,-5.2) -- ++(-8,0) node[midway, above]{\small \{EncryptedExtensions\}};
  \draw[->, blue] (server)++(0,-6.2) -- ++(-8,0) node[midway, above]{\small \{Certificate + CertificateVerify\}};
  \draw[->, blue] (server)++(0,-7.2) -- ++(-8,0) node[midway, above]{\small \{Finished\}};
  \draw[->, blue] (client)++(0,-8.2) -- ++(8,0) node[midway, above]{\small \{Finished\}};
  \draw[->, red] (client)++(0,-9.2) -- ++(8,0) node[midway, above]{\small HTTP GET /customers};
  \draw[->, red] (server)++(0,-10.0) -- ++(-8,0) node[midway, above]{\small HTTP 200 OK};
  \draw[decorate, decoration={brace, amplitude=5pt, mirror}] (-2,-0.3) -- (-2,-2.3) node[midway, left=8pt, font=\footnotesize] {TCP};
  \draw[decorate, decoration={brace, amplitude=5pt, mirror}] (-2,-3.0) -- (-2,-8.4) node[midway, left=8pt, font=\footnotesize] {TLS 1.3};
  \draw[decorate, decoration={brace, amplitude=5pt, mirror}] (-2,-9.0) -- (-2,-10.2) node[midway, left=8pt, font=\footnotesize] {App};
  \draw[decorate, decoration={brace, amplitude=3pt, mirror}, gray, dashed] (-2.8,-2.3) -- (-2.8,-3.0) node[midway, left=6pt, font=\scriptsize, text=gray] {TCP$\to$TLS};
  \draw[decorate, decoration={brace, amplitude=3pt, mirror}, gray, dashed] (-2.8,-8.4) -- (-2.8,-9.0) node[midway, left=6pt, font=\scriptsize, text=gray] {TLS$\to$App};
\end{tikzpicture}%
}%
\caption{TLS~1.3 full handshake over TCP with 1-RTT key exchange. Note: this study focuses on the 1-RTT full handshake mode (no 0-RTT early data, no session resumption) to measure the unmitigated impact of PQC on initial connection establishment.}
\label{fig:tls13_handshake}
\end{figure}

\subsection{TLS 1.3}
\label{sec:TLS}
TLS~1.3 (RFC~8446) is the latest version of the Transport Layer Security protocol, designed to provide confidentiality, integrity, and authentication for Internet communications. Compared to its predecessors, TLS~1.3 reduces the handshake to a single round trip (1-RTT), removes legacy cipher suites, and mandates forward secrecy through ephemeral key exchange. In a typical TLS~1.3 full handshake, the connection proceeds through three main protocol phases: TCP establishment, TLS negotiation, and application data exchange, each of which can be further decomposed into sub-phases. Figure~\ref{fig:tls13_handshake} illustrates this decomposition, showing the five layers defined in this study: the TCP handshake, the intermediate TCP-to-TLS delay, the TLS handshake itself, the intermediate TLS-to-Application delay, and the application response.

The rationale for decomposing the end-to-end latency into these five distinct layers is that each protocol phase is affected differently by PQC. Specifically: (i)~the \textbf{TCP handshake} (SYN to SYN-ACK) serves as a control layer---it should be independent of the cryptographic algorithm and any variation here indicates system-level effects; (ii)~the \textbf{TCP-to-TLS delay} (SYN-ACK to ClientHello) captures the client-side cost of TLS context initialization and key material generation, which is where PQC is expected to have the largest impact; (iii)~the \textbf{TLS handshake exchange} (ClientHello$\to$Finished) reflects the on-the-wire cost of key exchange, certificate transmission, and cryptographic verification; (iv)~the \textbf{TLS-to-Application delay} (Finished$\to$ HTTP~GET) isolates any residual overhead after secure channel establishment; and (v)~the \textbf{Application response} (HTTP~GET$\to$HTTP~200~OK) measures the client-to-backend response time, which should be algorithm-independent.

Note that the \textbf{TCP-to-TLS delay} is not a networking layer in the traditional protocol-stack sense. During this interval the client initializes the TLS context, generates the ephemeral key-exchange material for every group offered in the ClientHello, and serializes the ClientHello. This distinction is essential for interpreting the results: when we later describe the TLS handshake exchange as \emph{algorithm-neutral}, we refer specifically to the ClientHello$\to$Finished message exchange, not to the ClientHello \emph{construction} cost, which is measured separately in the TCP-to-TLS delay.

By analyzing each layer separately, we can precisely attribute the PQC overhead to the specific protocol phase where it originates, rather than observing only the aggregate effect on end-to-end latency.

\subsection{PQC Deployment in TLS}

The integration of post-quantum algorithms into production TLS stacks has progressed rapidly. Google Chrome enabled hybrid post-quantum key exchange by default in Chrome~124 (April 2024)~\cite{chrome2024}. Cloudflare reported large-scale deployment and operational considerations for hybrid TLS in production~\cite{cloudflare2023}. Amazon Web Services published production support updates for post-quantum TLS in AWS KMS and s2n-tls~\cite{awskms2019}. On the open-source side, the OQS project~\cite{oqs2024} provides liboqs and an OpenSSL provider that enables PQC algorithm negotiation in standard TLS stacks---the same toolchain used in this study.

Paquin et al.~\cite{paquin2020} conducted systematic benchmarks of PQC algorithms integrated into the OQS-OpenSSL fork, measuring handshake completion times and throughput for various KEM and signature combinations. Their work provides valuable computational baselines but evaluates aggregate handshake times without per-layer decomposition and predates the final NIST standards.

\subsection{Performance Studies and contribution of this work}

Several performance studies have addressed the impact of post-quantum algorithms on TLS. Sikeridis et al.~\cite{iacr2020} focus on post-quantum digital signatures and provide a comparative analysis of TLS authentication overhead. Similarly, Raabi et al.~\cite{mdpi2025} evaluate post-quantum signature schemes in isolation.

Sosnowski et al.~\cite{acm2024} compare TLS handshake latency across several pre- and post-quantum algorithms, reporting aggregate handshake times without per-layer decomposition. Kampanakis et al.~\cite{iacr2024} evaluate the Time-To-Last-Byte (TTLB) between P-256 and post-quantum algorithms using real-world connection data. Zheng et al.~\cite{springer2024} demonstrate that ML-KEM can achieve competitive performance despite larger key material. Liu et al.~\cite{liu2025} report on the PQC migration of a physical 5G testbed using ML-KEM~768 (hybrid with x25519) for TLS, IPSec, and SUCI; their evaluation shows minimal performance degradation (throughput drop ${<}$4.3\%, latency variation within $\pm$0.3\,ms), consistent with our finding that PQC overhead is concentrated in connection establishment and negligible at steady state.

In contrast to these works, our study (i)~decomposes the transaction into five protocol layers rather than measuring only end-to-end latency, (ii)~compares classical, hybrid, and pure PQC configurations side by side, and (iii)~demonstrates that the TLS handshake \emph{exchange} (ClientHello$\to$Finished) is effectively \emph{algorithm-neutral}---with the algorithm-sensitive cost isolated to ClientHello construction in the TCP-to-TLS layer---with all PQC configurations showing negligible effect sizes (Glass's $\Delta<0.33$)~\cite{glass1976,cohen1988}---a granularity of analysis not reported in prior literature. Note that hybrid key exchange (e.g., x25519\_MLKEM768) follows the IETF design draft~\cite{ietf_hybrid} and an active ECDHE-MLKEM specification track~\cite{ietf_ecdhe_mlkem}, and is not yet a finalized RFC, unlike the pure ML-KEM algorithms standardized in FIPS~203~\cite{nist_mlkem}.

To make the contribution boundaries explicit, Table~\ref{tab:related_work_comparison} positions prior studies against this work along four objective, verifiable criteria: scope (key exchange vs.\ signature), metric granularity (aggregate vs.\ per-layer), test methodology, and whether backend response size is varied as an independent variable.

\begin{table}[H]
\centering
\caption{Feature comparison with related work along objective criteria.}
\label{tab:related_work_comparison}
\resizebox{\textwidth}{!}{%
\begin{tabular}{lcccc}
\hline
\textbf{Study} & \textbf{Scope} & \textbf{Granularity} & \textbf{Test model} & \textbf{Response var.} \\
\hline
Paquin et al.~(2020)~\cite{paquin2020} & KEx + Sig & Aggregate & Microbenchmark & --- \\
Sikeridis et al.~(2020)~\cite{iacr2020} & Signature & Aggregate & Emulated & --- \\
Raabi et al.~(2023)~\cite{mdpi2025} & Signature & Aggregate & Emulated & --- \\
Sosnowski et al.~(2024)~\cite{acm2024} & KEx & Aggregate & Internet-scale & --- \\
Kampanakis et al.~(2024)~\cite{iacr2024} & KEx & Aggregate & Real-world & --- \\
Zheng et al.~(2024)~\cite{springer2024} & KEx & Aggregate & Emulated & --- \\
Liu et al.~(2025)~\cite{liu2025} & KEx + IPSec & Aggregate & Physical 5G & --- \\
\textbf{This work} & \textbf{KEx} & \textbf{5-layer} & \textbf{Emulated (100\,TPS)} & $\checkmark$ \\
\hline
\end{tabular}%
}
\end{table}

\section{Methodology and Experimental Setup}

\subsection{Objective}
The objective of this study is to evaluate the performance impact of integrating post-quantum cryptography (PQC) into the TLS~1.3 protocol. The experiments focus on handshake latency, message size, and connection throughput under realistic network conditions, comparing classical, hybrid, and post-quantum configurations.

\subsection{Environment Configuration}
\medskip
{\sloppy\textbf{Server side (Nginx reverse proxy).}
 To simulate the typical configuration in several production environments, we implemented a TLS-terminating endpoint using \textbf{Nginx~1.27.3}, compiled from source against a custom build of \textbf{OpenSSL~3.4.0} integrated with \textbf{liboqs~0.12.0} and the \textbf{OQS provider~0.8.0}. The base image is Alpine~3.21. Server certificates are signed with \textbf{RSA-2048} (\texttt{rsa:2048}). The supported key exchange groups are configured in the OpenSSL provider configuration and include: x25519,\allowbreak\ x448,\allowbreak\ prime256v1,\allowbreak\ secp384r1,\allowbreak\ mlkem512,\allowbreak\ mlkem768,\allowbreak\ mlkem1024,\allowbreak\ X25519MLKEM512,\allowbreak\ X25519MLKEM768,\allowbreak\ and SecP256r1MLKEM768. The system OpenSSL libraries were replaced by the custom build to avoid conflicts.\par}

\medskip
\textbf{Client side (Keysight CyPerf).} Traffic generation was performed using \textbf{Keysight CyPerf}~\cite{keysight2024}, a high-performance stateful traffic generator that emulates high-scale HTTPS clients and servers (with up to hundreds of thousands of TLS handshakes per second). CyPerf agents have built-in native support for ML-KEM key exchange groups, enabling negotiation of both classical and post-quantum algorithms during the TLS~1.3 handshake. The client workloads consist of concurrent virtual users executing stateful HTTP transactions at a sustained rate of 100~TPS to emulate realistic service load (well below the traffic generation tool's capability).

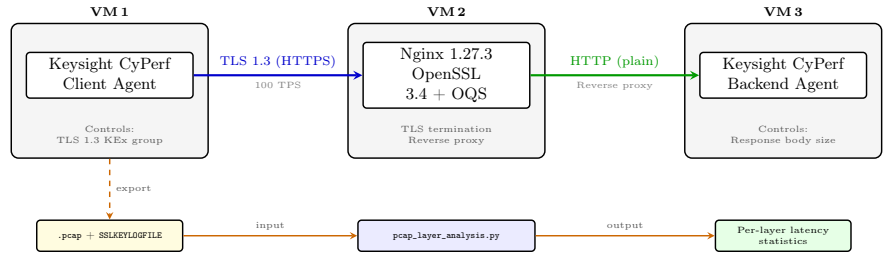
\begin{figure}[H]
\centering
\resizebox{0.95\textwidth}{!}{%
\begin{tikzpicture}[
  >=stealth, thick,
  vm/.style={draw, rounded corners=4pt, minimum width=3.8cm, minimum height=2.6cm, fill=gray!8},
  comp/.style={draw, rounded corners=2pt, fill=white, font=\footnotesize, text width=3cm, align=center, minimum height=0.7cm},
  lbl/.style={font=\scriptsize\bfseries, anchor=south},
]
  \node[vm] (vm1) at (0,0) {};
  \node[lbl] at (vm1.north) {VM\,1};
  \node[comp] (agent1) at (0,0.3) {Keysight CyPerf\\Client Agent};
  \node[font=\tiny, text=gray!70!black, anchor=north, text width=3cm, align=center] at (0,-0.55) {Controls:\\TLS 1.3 KEx group};

  \node[vm] (vm2) at (6.5,0) {};
  \node[lbl] at (vm2.north) {VM\,2};
  \node[comp] (nginx) at (6.5,0.3) {Nginx 1.27.3\\OpenSSL 3.4 + OQS};
  \node[font=\tiny, text=gray!70!black, anchor=north, text width=3cm, align=center] at (6.5,-0.55) {TLS termination\\Reverse proxy};

  \node[vm] (vm3) at (13,0) {};
  \node[lbl] at (vm3.north) {VM\,3};
  \node[comp] (agent2) at (13,0.3) {Keysight CyPerf\\Backend Agent};
  \node[font=\tiny, text=gray!70!black, anchor=north, text width=3cm, align=center] at (13,-0.55) {Controls:\\Response body size};

  \draw[->, blue!80!black, line width=1.2pt] (agent1.east) -- (nginx.west)
    node[midway, above, font=\scriptsize] {TLS 1.3 (HTTPS)}
    node[midway, below, font=\tiny, text=gray] {100 TPS};
  \draw[->, green!60!black, line width=1.2pt] (nginx.east) -- (agent2.west)
    node[midway, above, font=\scriptsize] {HTTP (plain)}
    node[midway, below, font=\tiny, text=gray] {Reverse proxy};

  \node[draw, rounded corners=2pt, fill=yellow!15, font=\tiny, text width=2.6cm, align=center, minimum height=0.6cm]
    (pcap) at (0,-2.8) {\texttt{.pcap} + \texttt{SSLKEYLOGFILE}};
  \draw[->, orange!80!black, line width=0.8pt, dashed] (vm1.south) -- (pcap.north)
    node[midway, right, font=\tiny, text=gray!70!black] {export};

  \node[draw, rounded corners=2pt, fill=blue!8, font=\tiny, text width=3.2cm, align=center, minimum height=0.6cm]
    (script) at (6.5,-2.8) {\texttt{pcap\_layer\_analysis.py}};
  \draw[->, orange!80!black, line width=0.8pt] (pcap.east) -- (script.west)
    node[midway, above, font=\tiny, text=gray!70!black] {input};

  \node[draw, rounded corners=2pt, fill=green!10, font=\tiny, text width=2.4cm, align=center, minimum height=0.6cm]
    (results) at (13,-2.8) {Per-layer latency\\statistics};
  \draw[->, orange!80!black, line width=0.8pt] (script.east) -- (results.west)
    node[midway, above, font=\tiny, text=gray!70!black] {output};
\end{tikzpicture}%
}%
\caption{Test bed architecture. Three VMs connected via local networking: the CyPerf client agent injects HTTPS traffic at 100~TPS towards Nginx (TLS termination with OQS), which proxies plain HTTP to the CyPerf backend agent. The key exchange group and response body size are independently controlled by CyPerf. The captured \texttt{pcap} files and TLS key logs are processed by the data-reduction script to extract per-layer latency statistics.}
\label{fig:testbed}
\end{figure}

\subsection{Metrics and Data Collection}
\label{sec:metrics}
Performance evaluation is organized into three metric families:

\medskip
\textbf{Per-layer latency metrics.} For each of the five protocol layers described in Section~\ref{sec:TLS}, the data-reduction tool computes the following descriptive statistics from all valid TCP streams in a capture: arithmetic mean, median (p50), p90, p95, and p99 percentiles, minimum, maximum, and standard deviation (SD). Additionally, the \textbf{total end-to-end time (E2E)}, defined as the per-connection wall-clock interval from the TCP~SYN to the final HTTP~200~OK, is computed per connection and summarized with its mean and SD. The E2E time is obtained directly from the per-connection boundary timestamps (not by summing per-layer percentiles, which are not additive). 

\medskip
\textbf{Message size and key material metrics.} For each algorithmic configuration, the \texttt{ClientHello} length (bytes) and \texttt{ServerHello} length (bytes) are recorded, as well as the client key exchange (key\_share) payload size. The \texttt{key\_share} payload size is deterministic for a given key exchange group, as it is fixed by the algorithm specification~\cite{nist_mlkem}. This allows direct quantification of the bandwidth overhead introduced by PQC key material.

\medskip
\textbf{Resource utilization metrics.} CPU utilization is recorded separately for client (CyPerf agent) and server (Nginx) VMs, broken down into \emph{System}, \emph{User}, and \emph{Idle} components. Network throughput (Mb/s) is captured on the server-facing interface (eth1~recv). These metrics enable correlation between cryptographic workload and system resource consumption.

\medskip
\textbf{Test scenarios.} Three backend response configurations are evaluated per algorithm: (i)~\emph{direct} (minimal response body), (ii)~\emph{4\,KB response body}, and (iii)~\emph{40\,KB response body}. Here the \emph{response body size} refers to the length of the HTTP payload returned by the backend server (and relayed by the Nginx reverse proxy) to the client in the application-data TLS records that follow handshake completion that is, the bytes delivered after the secure channel is established, not part of the handshake itself. This variation isolates the cryptographic overhead from the application-layer transfer cost.

Each configuration is executed multiple times to ensure statistical reliability.

\medskip
\textbf{Effect-size criterion.} To assess whether observed latency differences are practically meaningful, we compute Glass's~$\Delta$~\cite{glass1976}: the difference between the PQC and x25519 means, divided by the standard deviation (SD) of the x25519 baseline, which represents the inherent measurement variability without PQC influence. We interpret~$|\Delta|$ following Cohen's conventional thresholds~\cite{cohen1988}: $|\Delta|<0.2$ (negligible), $0.2$--$0.8$ (small to medium), and $|\Delta|>0.8$ (large).

\subsection{Data Reduction from Packet Captures}
{\sloppy To perform the layer-by-layer latency decomposition, full packet captures (\texttt{pcap} files) were recorded during each load test using \texttt{CyPerf}. A key requirement of this methodology is that, for \emph{each} performance-test run, TLS key material (session/master keys) is exported through \texttt{SSLKEYLOGFILE}; without this step, packet-level decryption and reliable request-level correlation across TCP/TLS/HTTP boundaries is not possible. CyPerf is able to export the \texttt{SSLKEYLOGFILE} as well, together with the associated packet capture. These captures were then processed by a custom Python script pcap\_layer\_analysis~\cite{pcap_layer_analysis} that decrypts and parses each TCP stream to extract per-connection timestamps at the following protocol boundaries: SYN, SYN-ACK, ClientHello, TLS Finished, HTTP GET, and HTTP 200 OK. For each connection, the tool computes the five inter-layer deltas and aggregates them into percentile-based statistics (p50, p90, p95, p99) across all streams in the capture file, enabling the statistical analysis presented in this paper. The tool supports parallel processing of large captures by splitting them into stream batches via \texttt{tshark}. The full repository and the script used for this step are publicly available~\cite{pcap_layer_analysis}.\par}

\subsection{Algorithmic Configurations}
The experiments compare several algorithmic combinations for TLS~1.3 key exchange. In all configurations, the server authentication uses \textbf{rsa:2048} certificates, generated by the OQS-enabled OpenSSL build. Since the signature algorithm is held constant across all tests, the measured latency differences are attributable exclusively to the key exchange algorithm. The key exchange groups tested are:
\begin{itemize}
  \item \textbf{Classical}: x25519 (Curve25519 ECDH).
  \item \textbf{Hybrid}: x25519\_MLKEM512, x25519\_MLKEM768 (combining classical ECDH with ML-KEM as defined in the IETF hybrid key exchange draft~\cite{ietf_hybrid}).
  \item \textbf{Pure Post-Quantum}: MLKEM512, MLKEM1024 (ML-KEM only, per FIPS~203~\cite{nist_mlkem}).
  \item Additional tests varying the digital signature algorithm (e.g., ECDSA vs.\ ML-DSA vs.\ SLH-DSA) to isolate signature overhead are planned as future work.
\end{itemize}

\subsection{Monitoring and Visualization}
All metrics are exported through \textbf{Prometheus} and visualized using \textbf{Grafana} and \textbf{CyPerf}, enabling percentile-based analysis of TLS handshake and data transfer times.  
This monitoring infrastructure supports long-duration experiments and simplifies the identification of bottlenecks in PQC-enabled TLS implementations.

\subsection{Reproducibility and Limitations}
To ensure reproducibility, the entire experimental setup is defined through containerized deployment manifests and Dockerfiles stored in a public Git repository.  
All containers are pinned to specific base images and software versions, including:
{\sloppy
\begin{itemize}
  \item \textbf{Base image}: Alpine~3.21.
  \item \textbf{OpenSSL}: 3.4.0 (built from source from the official repository).
  \item \textbf{liboqs}: 0.12.0 (Open Quantum Safe library for PQC primitives).
  \item \textbf{OQS provider}: 0.8.0 (OpenSSL provider enabling PQC algorithms).
  \item \textbf{Nginx}: 1.27.3, compiled with \texttt{--with-openssl} pointing to the custom OpenSSL build.
  \item \textbf{Server certificate signature}: rsa-2048 (\texttt{rsa:2048}).
  \item \textbf{Keysight CyPerf}: with standard, native ML-KEM support.
\end{itemize}
\par}

The main limitations of this study are:
\begin{itemize}
  \item The use of virtualized networking may slightly reduce absolute throughput compared to bare-metal environments.
  \item Only single-node Docker deployments were evaluated; distributed scenarios with inter-node latency were left for future work.
  \item All tests employ TLS~1.3 \textbf{1-RTT full handshake} mode with fresh connections (no session resumption, no 0-RTT early data). While 0-RTT would allow sending application data (HTTP GET) in the first client flight---thus significantly reducing perceived latency---this mode was deliberately excluded to isolate and measure the full cryptographic overhead of each algorithm without amortization. Production deployments may achieve lower latencies through session resumption or 0-RTT, but the PQC-induced overhead measured here represents the unavoidable cost for initial connections.
  \item The packet capture was performed in software, hence additional software processing layers might introduce extra delays.
\end{itemize}

\section{Results and Discussion}

This section presents and discusses the experimental results obtained from the layer-by-layer analysis of TLS~1.3 connections under classical, hybrid, and post-quantum cryptographic configurations. Latency results are reported in milliseconds, with percentile-based metrics emphasized to mitigate the influence of sporadic outliers. The analysis is organized into the following subsections:

\begin{itemize}
  \item \textbf{Section~\ref{sec:results:lat_breakdown}} provides an overview of the per-layer latency decomposition, summarizing the full dataset in a single reference table and a stacked visualization.
  \item \textbf{Section~\ref{sec:results:TCP-Handshake}} analyzes the TCP handshake layer, confirming its independence from the cryptographic algorithm.
  \item \textbf{Section~\ref{sec:results:TCP-to-TLS}} examines the TCP-to-TLS delay, where the dominant PQC overhead is localized.
  \item \textbf{Section~\ref{sec:results:TLS-Handshake}} evaluates the TLS handshake exchange latency, demonstrating its algorithm-neutral behavior.
  \item \textbf{Section~\ref{sec:results:TLS-to-App_App}} covers post-handshake and application-layer latency, showing negligible PQC impact.
  \item \textbf{Section~\ref{sec:normalized_overhead}} introduces normalized metrics (Overhead Factor and Cryptographic Overhead Share) to isolate and quantify the PQC penalty.
  \item \textbf{Section~\ref{sec:payload_impact}} evaluates the effect of increasing the backend response payload from 4\,KB to 40\,KB.
  \item \textbf{Section~\ref{sec:cpu_net}} reports CPU and network utilization across configurations.
  \item \textbf{Section~\ref{sec:e2e}} consolidates the end-to-end perspective, combining per-layer results into aggregate connection times.
\end{itemize}

\subsection{Latency Breakdown by Protocol Layer}
\label{sec:results:lat_breakdown}

Table~\ref{tab:latency_summary} summarizes the observed latency percentiles for the different protocol layers under the backend 4\,KB response body, along with 40\,KB response tests for x25519 and x25519\_MLKEM768 to assess PQC impact on larger payloads (analyzed in Section~\ref{sec:payload_impact}). Each layer is defined by its start and end timestamps in the TLS connection lifecycle. Five representative configurations are compared: 

\begin{table}[H]
\centering
\caption{Latency percentiles and standard deviation per protocol layer. Rows labeled (4\,KB) and (40\,KB) refer to the backend response body size. Comparing x25519 (classical), x25519\_MLKEM512 (hybrid), x25519\_MLKEM768 (hybrid), MLKEM512 (pure PQC), and MLKEM1024 (pure PQC). The 40\,KB scenario is included for x25519 and x25519\_MLKEM768 to support the analysis in Section~\ref{sec:payload_impact}.}
\label{tab:latency_summary}
\begin{tabular}{lcccc}
\hline
\textbf{Layer (Timestamp Start $\to$ End)} & \textbf{p50 (ms)} & \textbf{p95 (ms)} & \textbf{p99 (ms)} & \textbf{SD (ms)} \\
\hline
\multicolumn{5}{l}{\textit{TCP handshake (SYN $\to$ SYN-ACK)}} \\
\quad x25519 (4\,KB) & 0.360 & 0.694 & 0.957 & 0.248 \\
\quad x25519\_MLKEM512 (4\,KB) & 0.390 & 3.060 & 4.147 & 0.999 \\
\quad x25519\_MLKEM768 (4\,KB) & 0.402 & 2.720 & 4.139 & 1.016 \\
\quad MLKEM512 (4\,KB) & 0.381 & 2.882 & 3.892 & 0.969 \\
\quad MLKEM1024 (4\,KB) & 0.393 & 2.397 & 3.901 & 0.915 \\
\quad x25519 (40\,KB) & 0.401 & 0.781 & 1.251 & 16.993 \\
\quad x25519\_MLKEM768 (40\,KB) & 0.417 & 3.386 & 5.048 & 10.163 \\
\hline
\multicolumn{5}{l}{\textit{TCP-to-TLS delay (SYN-ACK $\to$ ClientHello)}} \\
\quad x25519 (4\,KB) & 0.294 & 0.635 & 0.800 & 0.194 \\
\quad x25519\_MLKEM512 (4\,KB) & 1.866 & 4.122 & 5.664 & 1.196 \\
\quad x25519\_MLKEM768 (4\,KB) & 1.903 & 3.900 & 5.534 & 1.411 \\
\quad MLKEM512 (4\,KB) & 1.726 & 3.577 & 5.043 & 1.065 \\
\quad MLKEM1024 (4\,KB) & 1.772 & 3.608 & 5.088 & 1.044 \\
\quad x25519 (40\,KB) & 0.289 & 0.702 & 0.901 & 0.232 \\
\quad x25519\_MLKEM768 (40\,KB) & 1.915 & 4.097 & 5.598 & 1.421 \\
\hline
\multicolumn{5}{l}{\textit{TLS handshake (ClientHello $\to$ Finished)}} \\
\quad x25519 (4\,KB) & 5.547 & 10.903 & 12.697 & 2.893 \\
\quad x25519\_MLKEM512 (4\,KB) & 5.879 & 11.296 & 13.256 & 2.871 \\
\quad x25519\_MLKEM768 (4\,KB) & 6.495 & 11.692 & 13.650 & 2.982 \\
\quad MLKEM512 (4\,KB) & 5.253 & 9.999 & 12.010 & 3.798 \\
\quad MLKEM1024 (4\,KB) & 5.450 & 10.578 & 12.209 & 2.701 \\
\quad x25519 (40\,KB) & 5.791 & 10.886 & 12.242 & 2.827 \\
\quad x25519\_MLKEM768 (40\,KB) & 6.105 & 10.715 & 12.603 & 7.055 \\
\hline
\multicolumn{5}{l}{\textit{TLS-to-Application delay (Finished $\to$ HTTP GET)}} \\
\quad x25519 (4\,KB) & 0.526 & 1.227 & 1.912 & 0.389 \\
\quad x25519\_MLKEM512 (4\,KB) & 0.991 & 2.363 & 3.479 & 0.747 \\
\quad x25519\_MLKEM768 (4\,KB) & 1.004 & 2.576 & 3.730 & 0.976 \\
\quad MLKEM512 (4\,KB) & 0.897 & 2.605 & 3.647 & 1.741 \\
\quad MLKEM1024 (4\,KB) & 0.950 & 2.536 & 3.554 & 0.635 \\
\quad x25519 (40\,KB) & 0.559 & 1.272 & 2.061 & 0.415 \\
\quad x25519\_MLKEM768 (40\,KB) & 0.989 & 2.546 & 3.741 & 0.849 \\
\hline
\multicolumn{5}{l}{\textit{Application response (HTTP GET $\to$ HTTP 200 OK)}} \\
\quad x25519 (4\,KB) & 9.071 & 14.866 & 17.424 & 3.553 \\
\quad x25519\_MLKEM512 (4\,KB) & 8.880 & 14.925 & 17.510 & 3.544 \\
\quad x25519\_MLKEM768 (4\,KB) & 8.334 & 14.186 & 16.868 & 3.375 \\
\quad MLKEM512 (4\,KB) & 9.087 & 15.099 & 17.802 & 3.581 \\
\quad MLKEM1024 (4\,KB) & 8.838 & 14.442 & 17.026 & 3.451 \\
\quad x25519 (40\,KB) & 12.032 & 18.935 & 22.169 & 25.467 \\
\quad x25519\_MLKEM768 (40\,KB) & 11.166 & 18.479 & 21.721 & 16.792 \\
\hline
\end{tabular}
\end{table}

\begin{figure}[H]
\centering
\begin{minipage}[b]{0.49\textwidth}
\centering
\begin{tikzpicture}
\begin{axis}[
    ybar stacked, bar width=7pt,
    width=\textwidth, height=6cm,
    title={\footnotesize (a) p50 (median latency)},
    ylabel={\footnotesize Latency (ms)},
    symbolic x coords={x25519,Hyb-512,Hyb-768,ML-512,ML-1024},
    xtick=data, x tick label style={rotate=35, anchor=east, font=\tiny},
    ymin=0, ymax=21,
    legend to name=sharedlegend,
    legend style={font=\tiny, legend columns=5},
    tick label style={font=\tiny},
    ytick={0,5,...,20},
]
\addplot+[fill=blue!80, draw=blue!90!black] coordinates {(x25519,0.360) (Hyb-512,0.390) (Hyb-768,0.402) (ML-512,0.381) (ML-1024,0.393)};
\addplot+[fill=orange!85!black, draw=orange!95!black] coordinates {(x25519,0.294) (Hyb-512,1.866) (Hyb-768,1.903) (ML-512,1.726) (ML-1024,1.772)};
\addplot+[fill=green!70!black, draw=green!90!black] coordinates {(x25519,5.547) (Hyb-512,5.879) (Hyb-768,6.495) (ML-512,5.253) (ML-1024,5.450)};
\addplot+[fill=red!80, draw=red!90!black] coordinates {(x25519,0.526) (Hyb-512,0.991) (Hyb-768,1.004) (ML-512,0.897) (ML-1024,0.950)};
\addplot+[fill=violet!80!black, draw=violet!95!black] coordinates {(x25519,9.071) (Hyb-512,8.880) (Hyb-768,8.334) (ML-512,9.087) (ML-1024,8.838)};
\legend{TCP Handshake, TCP to TLS Delay, TLS Handshake, TLS to App Delay, Application}
\end{axis}
\end{tikzpicture}
\end{minipage}
\hfill
\begin{minipage}[b]{0.49\textwidth}
\centering
\begin{tikzpicture}
\begin{axis}[
    ybar stacked, bar width=7pt,
    width=\textwidth, height=6cm,
    title={\footnotesize (b) p95 (tail latency)},
    ylabel={\footnotesize Latency (ms)},
    symbolic x coords={x25519,Hyb-512,Hyb-768,ML-512,ML-1024},
    xtick=data, x tick label style={rotate=35, anchor=east, font=\tiny},
    ymin=0, ymax=38,
    tick label style={font=\tiny},
    ytick={0,5,10,...,35},
]
\addplot+[fill=blue!80, draw=blue!90!black] coordinates {(x25519,0.694) (Hyb-512,3.060) (Hyb-768,2.720) (ML-512,2.882) (ML-1024,2.397)};
\addplot+[fill=orange!85!black, draw=orange!95!black] coordinates {(x25519,0.635) (Hyb-512,4.122) (Hyb-768,3.900) (ML-512,3.577) (ML-1024,3.608)};
\addplot+[fill=green!70!black, draw=green!90!black] coordinates {(x25519,10.903) (Hyb-512,11.296) (Hyb-768,11.692) (ML-512,9.999) (ML-1024,10.578)};
\addplot+[fill=red!80, draw=red!90!black] coordinates {(x25519,1.227) (Hyb-512,2.363) (Hyb-768,2.576) (ML-512,2.605) (ML-1024,2.536)};
\addplot+[fill=violet!80!black, draw=violet!95!black] coordinates {(x25519,14.866) (Hyb-512,14.925) (Hyb-768,14.186) (ML-512,15.099) (ML-1024,14.442)};
\end{axis}
\end{tikzpicture}
\end{minipage}
\par\vspace{4pt}
\ref{sharedlegend}
\caption{End-to-end latency decomposition by protocol layer (backend 4\,KB). Left: p50 (median). Right: p95 (tail). Abbreviations: Hyb-512\,=\,x25519\_MLKEM512, Hyb-768\,=\,x25519\_MLKEM768, ML-512\,=\,MLKEM512, ML-1024\,=\,MLKEM1024.}
\label{fig:stacked_layers}
\end{figure}
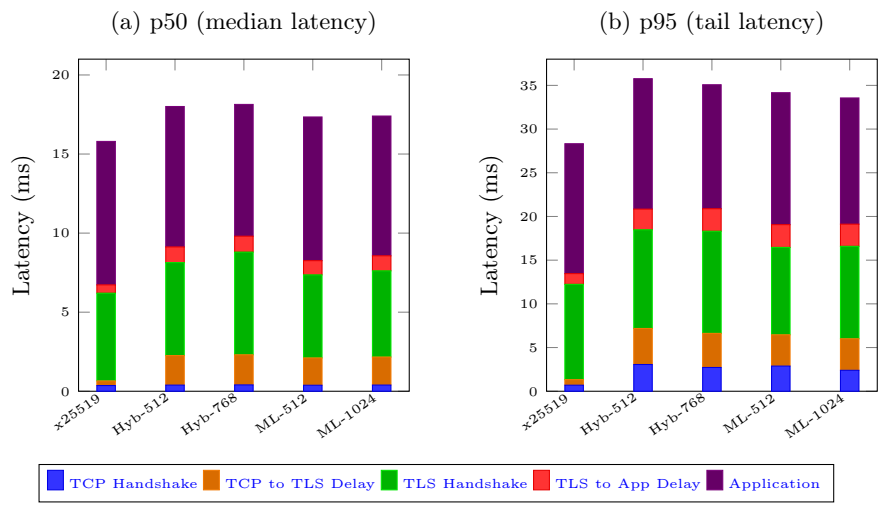

Figure~\ref{fig:stacked_layers} provides a visual decomposition of the end-to-end latency by protocol layer at both the median (p50) and tail (p95) percentiles for each algorithm under the 4\,KB backend scenario.

 Comparing (a) and (b) reveals that the TCP and TCP$\to$TLS layers grow disproportionately at the tail under PQC, while TLS and application layers scale proportionally. Additional tests with a 40\,KB backend response body were also conducted for x25519 and x25519\_MLKEM768; the corresponding analysis is presented in Section~\ref{sec:payload_impact}
\subsection{TCP Handshake Latency}
\label{sec:results:TCP-Handshake}

The TCP handshake latency results in Table~\ref{tab:latency_summary} show no statistically significant dependency on the cryptographic algorithm at the median: values range from 0.360 to 0.402\,ms across all configurations, a maximum difference of 0.042\,ms ($\Delta=0.17$, negligible per Cohen's thresholds; Section~\ref{sec:metrics}). This confirms that the TCP layer is genuinely algorithm-independent. However, at p95, tail latencies increase under PQC configurations (up to 3.06\,ms for x25519\_MLKEM512 vs.\ 0.694\,ms for x25519)---an effect attributable to higher system load rather than cryptographic overhead, since the TCP layer precedes any key exchange computation.

\subsection{Client-Side TLS Initialization Cost}
\label{sec:results:TCP-to-TLS}

The TCP-to-TLS delay (SYN-ACK to ClientHello) captures the cost of TLS context initialization and ClientHello construction. This layer exhibits the most pronounced post-quantum impact, aligned with real-world expectations, with median latency increasing from approximately 0.3\,ms for x25519 to around 1.8--1.9\,ms for both hybrid and pure post-quantum configurations (see Table~\ref{tab:latency_summary}).

This increase is consistent across MLKEM variants and independent of backend response size, indicating that the dominant cost arises from client-side initialization and serialization.

Because this cost is incurred entirely before the first byte is placed on the wire, it is also a prime candidate for optimization. A TLS library could \emph{precompute} the ephemeral ClientHello key material ahead of connection establishment---for example, by maintaining a small pool of ready-to-use key shares for the offered groups---moving post-quantum key generation off the connection critical path. Such precomputation would render the TCP-to-TLS overhead factor effectively negligible without altering the on-the-wire protocol, since the heavy operation (ephemeral key-pair generation) is independent of the peer and can be performed in advance.

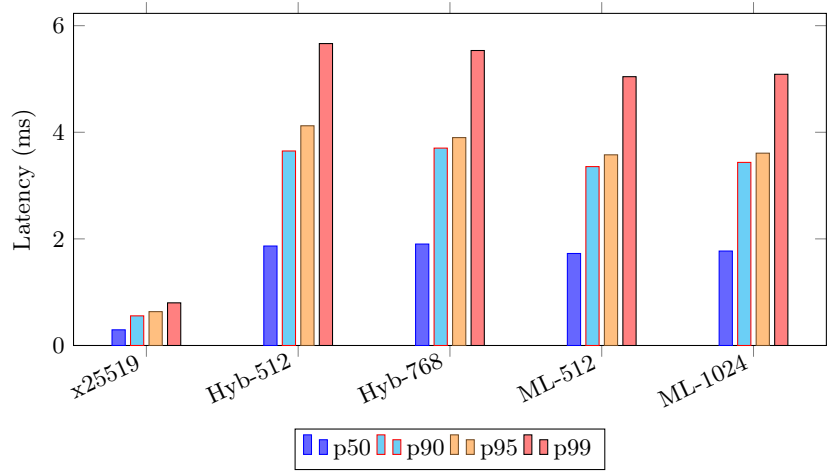
\begin{figure}[H]
\centering
\begin{tikzpicture}
\begin{axis}[
    ybar, bar width=5pt,
    width=0.95\textwidth, height=6cm,
    ylabel={Latency (ms)},
    symbolic x coords={x25519,Hyb-512,Hyb-768,ML-512,ML-1024},
    xtick=data, x tick label style={rotate=25, anchor=east, font=\footnotesize},
    legend style={at={(0.02,0.98)}, anchor=north west, font=\footnotesize},
    ymin=0, enlarge x limits=0.12,
    legend style={
        at={(0.5,-0.25)},
        anchor=north,
        font=\footnotesize,
    },
    legend columns=4,        
]
\addplot+[fill=blue!60] coordinates {(x25519,0.294) (Hyb-512,1.866) (Hyb-768,1.903) (ML-512,1.726) (ML-1024,1.772)};
\addplot+[fill=cyan!50] coordinates {(x25519,0.556) (Hyb-512,3.649) (Hyb-768,3.703) (ML-512,3.356) (ML-1024,3.435)};
\addplot+[fill=orange!50] coordinates {(x25519,0.635) (Hyb-512,4.122) (Hyb-768,3.900) (ML-512,3.577) (ML-1024,3.608)};
\addplot+[fill=red!50] coordinates {(x25519,0.800) (Hyb-512,5.664) (Hyb-768,5.534) (ML-512,5.043) (ML-1024,5.088)};
\legend{p50, p90, p95, p99}
\end{axis}
\end{tikzpicture}
\caption{TCP-to-TLS delay (SYN-ACK to ClientHello): full percentile profile per algorithm (backend 4\,KB). Abbreviations: Hyb-512\,=\,x25519\_MLKEM512, Hyb-768\,=\,x25519\_MLKEM768, ML-512\,=\,MLKEM512, ML-1024\,=\,MLKEM1024.}
\label{fig:tcp_tls_delay}
\end{figure}

Figure~\ref{fig:tcp_tls_delay} presents the full percentile profile (p50, p90, p95, p99) for this layer, revealing that the $6\times$ overhead at the median is preserved across all percentiles---a strong indicator that the latency shift is consistent rather than driven by outliers.

\subsection{TLS Handshake Latency}
\label{sec:results:TLS-Handshake}

{\sloppy The TLS handshake latency (ClientHello to Finished) reveals a nuanced picture. \textbf{Hybrid} configurations (x25519\_MLKEM512, x25519\_MLKEM768) show median values of 5.879\,ms and 6.495\,ms respectively, compared to 5.547\,ms for x25519. However, these differences (0.332--0.948\,ms) must be evaluated against the baseline standard deviation of 2.893\,ms: Glass's $\Delta=0.11$--$0.33$, well below Cohen's negligible threshold of~0.2 for the smaller hybrid and negligible-to-small for x25519\_MLKEM768 (Section~\ref{sec:metrics}), indicating that the effect is \emph{practically negligible}. Similarly, \textbf{pure ML-KEM} variants show slightly lower median latencies (MLKEM512: 5.253\,ms, MLKEM1024: 5.450\,ms), but the differences ($-$0.097 to $-$0.294\,ms, $|\Delta|=0.03$--$0.10$) are negligible and cannot be attributed to a genuine computational advantage. The observation that ML-KEM operations are inherently efficient---relying on parallel polynomial arithmetic over structured lattices rather than sequential scalar multiplication~\cite{bos2018,paquin2020,schwabe2024}---is well-documented in microbenchmarks, but at the TLS protocol level the differences are absorbed by protocol framing, network jitter, and system scheduling, making them indistinguishable from the baseline in our measurements.\par}

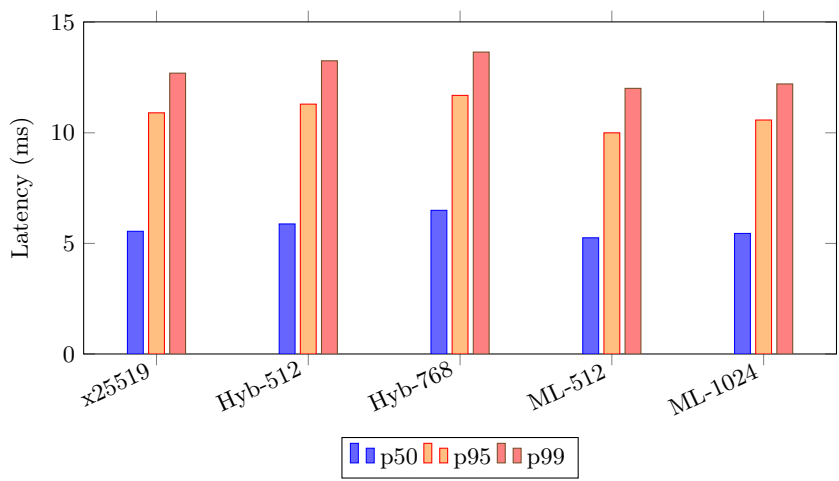
\begin{figure}[H]
\centering
\begin{tikzpicture}
\begin{axis}[
    ybar, bar width=6pt,
    width=0.95\textwidth, height=6cm,
    ylabel={Latency (ms)},
    symbolic x coords={x25519,Hyb-512,Hyb-768,ML-512,ML-1024},
    xtick=data, x tick label style={rotate=25, anchor=east, font=\footnotesize},
    legend style={at={(0.02,0.98)}, anchor=north west, font=\footnotesize},
    ymin=0, enlarge x limits=0.12,
    legend style={
        at={(0.5,-0.25)},
        anchor=north,
        font=\footnotesize,
    },
    legend columns=3,    
]
\addplot+[fill=blue!60] coordinates {(x25519,5.547) (Hyb-512,5.879) (Hyb-768,6.495) (ML-512,5.253) (ML-1024,5.450)};
\addplot+[fill=orange!50] coordinates {(x25519,10.903) (Hyb-512,11.296) (Hyb-768,11.692) (ML-512,9.999) (ML-1024,10.578)};
\addplot+[fill=red!50] coordinates {(x25519,12.697) (Hyb-512,13.256) (Hyb-768,13.650) (ML-512,12.010) (ML-1024,12.209)};
\legend{p50, p95, p99}

\end{axis}
\end{tikzpicture}
\caption{TLS handshake latency (ClientHello to Finished) per algorithm (backend 4\,KB). Abbreviations: Hyb-512\,=\,x25519\_MLKEM512, Hyb-768\,=\,x25519\_MLKEM768, ML-512\,=\,MLKEM512, ML-1024\,=\,MLKEM1024.}
\label{fig:tls_handshake}
\end{figure}

{\sloppy Figure~\ref{fig:tls_handshake} illustrates the TLS handshake latency for p50, p95, and p99 across all evaluated algorithms, graphically supporting the previous observation that the differences remain small and within measurement noise.\par}

\subsection{Post-Handshake and Application Latency}
\label{sec:results:TLS-to-App_App}

The TLS-to-application delay reflects residual effects immediately after secure channel establishment. Median latency increases from 0.526\,ms (x25519, SD\,=\,0.389\,ms) to 0.897--1.004\,ms for post-quantum configurations. Glass's $\Delta=0.95$--$1.23$, which Cohen's scale classifies as a large effect size (Section~\ref{sec:metrics}). However, the absolute magnitude is small (0.37--0.48\,ms), stable across configurations, and independent of backend response size, suggesting that while statistically detectable, the practical impact is limited.

Application response latency is dominated by end-to-end network path latency and response size. Backend responses with a 4\,KB body increase latency to approximately 8--9\,ms. Across all cryptographic algorithms, the maximum deviation from the x25519 baseline is 0.74\,ms (overall range 8.334--9.087\,ms, span 0.75\,ms), yielding $\Delta=0.21$---negligible to small per Cohen's thresholds (Section~\ref{sec:metrics}). This confirms that no practically meaningful application-level penalty attributable to post-quantum cryptography is observed. The effect of larger response payloads (40\,KB) on the PQC overhead is analyzed in Section~\ref{sec:payload_impact}.

\begin{figure}[H]
\centering
\begin{tikzpicture}
\begin{axis}[
    ybar, bar width=5pt,
    width=0.95\textwidth, height=6.5cm,
    ylabel={Latency (ms)},
    symbolic x coords={x25519,Hyb-512,Hyb-768,ML-512,ML-1024},
    xtick=data, x tick label style={rotate=25, anchor=east, font=\footnotesize},
    legend style={at={(0.5,-0.22)}, anchor=north, legend columns=2, font=\footnotesize},
    ymin=0, enlarge x limits=0.12,
]
\addplot+[fill=blue!80, draw=blue!90!black] coordinates {(x25519,1.883) (Hyb-512,2.030) (Hyb-768,2.054) (ML-512,1.804) (ML-1024,1.933)};
\addplot+[fill=orange!85!black, draw=orange!95!black] coordinates {(x25519,4.641) (Hyb-512,5.008) (Hyb-768,5.120) (ML-512,4.880) (ML-1024,5.039)};
\addplot+[fill=green!70!black, draw=green!90!black] coordinates {(x25519,9.071) (Hyb-512,8.880) (Hyb-768,8.334) (ML-512,9.087) (ML-1024,8.838)};
\addplot+[fill=red!80, draw=red!90!black] coordinates {(x25519,14.866) (Hyb-512,14.925) (Hyb-768,14.186) (ML-512,15.099) (ML-1024,14.442)};
\legend{Direct p50, Direct p95, Backend~4KB p50, Backend~4KB p95}
\end{axis}
\end{tikzpicture}
\caption{Application response latency per algorithm and scenario. Within each group, the near-identical bar heights confirm that PQC has no measurable impact at the application layer, at either percentile. Abbreviations: Hyb-512\,=\,x25519\_MLKEM512, Hyb-768\,=\,x25519\_MLKEM768, ML-512\,=\,MLKEM512, ML-1024\,=\,MLKEM1024.}
\label{fig:app_response}
\end{figure}
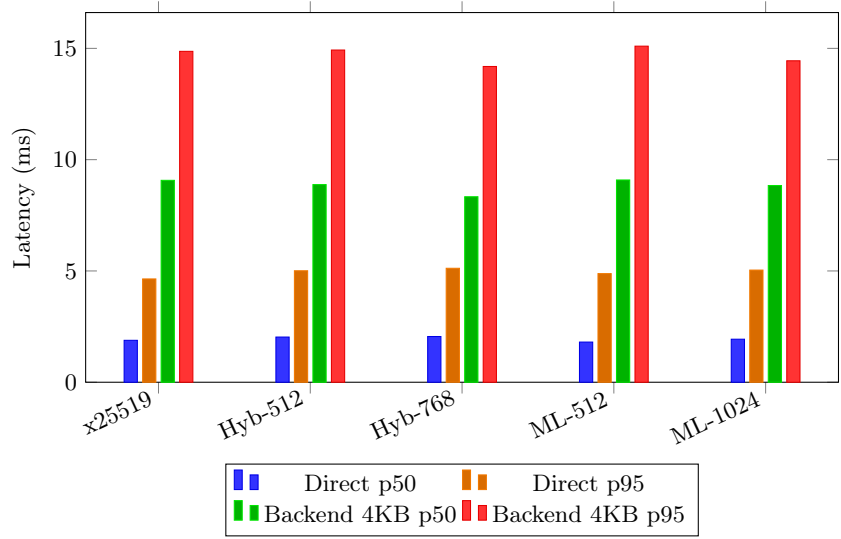

\subsection{Normalized Cryptographic Overhead}
\label{sec:normalized_overhead}

The preceding per-layer analysis reveals that only the TCP-to-TLS and TLS handshake layers exhibit measurable cryptographic sensitivity, while TCP, TLS-to-App, and application layers behave as algorithm-independent. To isolate and quantify this impact more precisely, we now normalize the overhead of these two crypto-sensitive layers relative to the classical baseline, providing a clearer measure of the actual PQC cost within the end-to-end transaction.

We define two complementary metrics. The \textbf{Overhead Factor (OF)} captures the per-layer slowdown relative to the classical baseline:
\begin{equation}
\text{OF}^{(l)} = \frac{L_{\text{PQC}}^{(l)}}{L_{\text{x25519}}^{(l)}}
\label{eq:of}
\end{equation}
where $L^{(l)}$ denotes the latency at a given percentile for protocol layer~$l$. An OF of~1.0 indicates no overhead; values below~1.0 indicate a speedup.

The \textbf{Cryptographic Overhead Share (COS)} aggregates the overhead of the two cryptographic-sensitive layers---TCP-to-TLS and TLS handshake---and expresses it as a fraction of the total end-to-end connection time:
\begin{equation}
\text{COS} = \frac{\displaystyle\sum_{l \,\in\, \{\text{TCP-to-TLS},\;\text{TLS}\}} \bigl(L_{\text{PQC}}^{(l)} - L_{\text{x25519}}^{(l)}\bigr)}{L_{\text{PQC}}^{(\text{e2e})}} \times 100\%
\label{eq:cos}
\end{equation}
COS quantifies what fraction of the total PQC connection time is attributable \emph{exclusively} to post-quantum cryptographic overhead, isolating it from protocol and application contributions. A COS of~0\% means no cryptographic penalty; higher values indicate a larger share of connection time consumed by PQC.

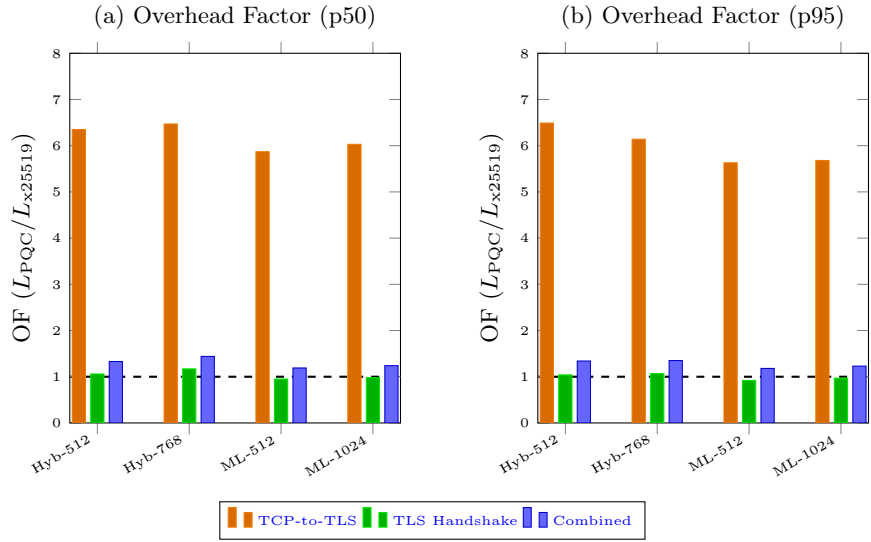
\begin{figure}[H]
\centering
\begin{minipage}[b]{0.49\textwidth}
\centering
\begin{tikzpicture}
\begin{axis}[
    ybar, bar width=5pt,
    width=\textwidth, height=6.5cm,
    title={\footnotesize (a) Overhead Factor (p50)},
    ylabel={\footnotesize OF ($L_{\text{PQC}} / L_{\text{x25519}}$)},
    symbolic x coords={Hyb-512,Hyb-768,ML-512,ML-1024},
    xtick=data, x tick label style={rotate=30, anchor=east, font=\tiny},
    ymin=0, ymax=8,
    legend to name=oflegend,
    legend style={font=\tiny, legend columns=3},
    tick label style={font=\tiny},
    ytick={0,1,...,8},
    extra y ticks={1},
    extra y tick style={grid=major, grid style={black, dashed, thick}},
    extra y tick labels={},
]
\addplot+[fill=orange!85!black, draw=orange!95!black] coordinates {(Hyb-512,6.35) (Hyb-768,6.47) (ML-512,5.87) (ML-1024,6.03)};
\addplot+[fill=green!70!black, draw=green!90!black] coordinates {(Hyb-512,1.06) (Hyb-768,1.17) (ML-512,0.95) (ML-1024,0.98)};
\addplot+[fill=blue!60, draw=blue!80!black] coordinates {(Hyb-512,1.33) (Hyb-768,1.44) (ML-512,1.19) (ML-1024,1.24)};
\legend{TCP-to-TLS, TLS Handshake, Combined}
\end{axis}
\end{tikzpicture}
\end{minipage}
\hfill
\begin{minipage}[b]{0.49\textwidth}
\centering
\begin{tikzpicture}
\begin{axis}[
    ybar, bar width=5pt,
    width=\textwidth, height=6.5cm,
    title={\footnotesize (b) Overhead Factor (p95)},
    ylabel={\footnotesize OF ($L_{\text{PQC}} / L_{\text{x25519}}$)},
    symbolic x coords={Hyb-512,Hyb-768,ML-512,ML-1024},
    xtick=data, x tick label style={rotate=30, anchor=east, font=\tiny},
    ymin=0, ymax=8,
    tick label style={font=\tiny},
    ytick={0,1,...,8},
    extra y ticks={1},
    extra y tick style={grid=major, grid style={black, dashed, thick}},
    extra y tick labels={},
]
\addplot+[fill=orange!85!black, draw=orange!95!black] coordinates {(Hyb-512,6.49) (Hyb-768,6.14) (ML-512,5.63) (ML-1024,5.68)};
\addplot+[fill=green!70!black, draw=green!90!black] coordinates {(Hyb-512,1.04) (Hyb-768,1.07) (ML-512,0.92) (ML-1024,0.97)};
\addplot+[fill=blue!60, draw=blue!80!black] coordinates {(Hyb-512,1.34) (Hyb-768,1.35) (ML-512,1.18) (ML-1024,1.23)};
\end{axis}
\end{tikzpicture}
\end{minipage}
\par\vspace{4pt}
\ref{oflegend}
\caption{Overhead Factor per algorithm for the two cryptographic-sensitive layers and their combination. The dashed line at OF\,=\,1 marks the x25519 baseline. Left: p50. Right: p95. The TCP-to-TLS layer dominates the overhead ($\sim$$6\times$, $\Delta\approx7.7$, far exceeding Cohen's large-effect threshold). $\text{OF}_{\text{TLS-HS}}$ values remain near 1.0 for all configurations; $\Delta=0.03$--$0.33$, negligible per Cohen's thresholds (Section~\ref{sec:metrics}).}
\label{fig:normalized_of}
\end{figure}

Table~\ref{tab:normalized_overhead} and Figure~\ref{fig:normalized_of} present these metrics for all PQC configurations at the median (p50) and tail (p95) percentiles.

\begin{table}[H]
\centering
\caption{Normalized cryptographic overhead per algorithm. OF: Overhead Factor (Eq.~\ref{eq:of}); COS: Cryptographic Overhead Share (Eq.~\ref{eq:cos}). Baseline x25519 has OF\,=\,1.00 and COS\,=\,0\% by definition.}
\label{tab:normalized_overhead}
\resizebox{\textwidth}{!}{%
\begin{tabular}{lccccc}
\hline
\textbf{Algorithm} & $\mathbf{OF_{\text{TCP-to-TLS}}}$ & $\mathbf{OF_{\text{TLS-HS}}}$ & $\mathbf{OF_{\text{Combined}}}$\textsuperscript{*} & \textbf{COS (\%)} \\
\hline
\multicolumn{5}{l}{\textit{p50 (median)}} \\
\quad x25519 & 1.00 & 1.00 & 1.00 & --- \\
\quad x25519\_MLKEM512 & 6.35 & 1.06 & 1.33 & 10.6 \\
\quad x25519\_MLKEM768 & 6.47 & 1.17 & 1.44 & 14.1 \\
\quad MLKEM512 & 5.87 & 0.95 & 1.19 & 6.6 \\
\quad MLKEM1024 & 6.03 & 0.98 & 1.24 & 7.9 \\
\hline
\multicolumn{5}{l}{\textit{p95 (tail latency)}} \\
\quad x25519 & 1.00 & 1.00 & 1.00 & --- \\
\quad x25519\_MLKEM512 & 6.49 & 1.04 & 1.34 & 10.8 \\
\quad x25519\_MLKEM768 & 6.14 & 1.07 & 1.35 & 11.6 \\
\quad MLKEM512 & 5.63 & 0.92 & 1.18 & 6.0 \\
\quad MLKEM1024 & 5.68 & 0.97 & 1.23 & 7.9 \\
\hline
\multicolumn{5}{l}{\footnotesize\textsuperscript{*}$\text{OF}_{\text{Combined}} = (L_{\text{TCP-to-TLS}}^{\text{PQC}} + L_{\text{TLS}}^{\text{PQC}}) / (L_{\text{TCP-to-TLS}}^{\text{x25519}} + L_{\text{TLS}}^{\text{x25519}})$} \\
\end{tabular}%
}
\end{table}

\medskip
\textbf{Key findings.} The TCP-to-TLS layer consistently exhibits an overhead factor ($\text{OF}_{\text{TCP-to-TLS}}$, Table~\ref{tab:normalized_overhead}) of approximately $6\times$ across all PQC variants and percentiles. This overhead is unambiguously significant: Glass's $\Delta\approx7.7$ (absolute difference $\sim$1.5\,ms vs.\ baseline SD of 0.194\,ms), far exceeding Cohen's large-effect threshold of~0.8 (Section~\ref{sec:metrics}), confirming that client-side key material generation is the dominant and systematic source of PQC latency. In contrast, the TLS handshake exchange overhead factor ($\text{OF}_{\text{TLS-HS}}$) remains close to~1.0: hybrid algorithms show a nominal increase (1.04--1.17$\times$) and pure ML-KEM variants show $\text{OF}_{\text{TLS-HS}}$ values nominally \emph{below}~1.0 (0.92--0.98$\times$). However, these TLS handshake differences correspond to $\Delta=0.03$--$0.33$ (absolute deltas of 0.1--0.9\,ms against a baseline SD of 2.893\,ms), well below Cohen's small-effect threshold of~0.5, meaning they are \emph{practically negligible}. The data is therefore consistent with the TLS handshake exchange being effectively \emph{algorithm-neutral} at this load level, rather than demonstrating a clear speedup or slowdown.

{\sloppy The COS metric reveals that, despite the large per-layer overhead in TCP-to-TLS, the cryptographic penalty accounts for only 6--14\% of the total end-to-end connection time. Hybrid configurations exhibit the highest COS (10.6--14.1\%), as they combine ECDH and KEM costs, whereas pure ML-KEM configurations achieve a lower COS (6.0--7.9\%). Note, however, that the COS difference between hybrid and pure ML-KEM configurations is driven primarily by the $\text{OF}_{\text{TLS-HS}}$ values, which---as established above---are not statistically distinguishable from 1.0. The robust conclusion is that the overall COS is moderate (below 15\%) for all PQC configurations tested.\par}

{\sloppy\textbf{On the relationship between computational cost and security level.} Although the $\text{OF}_{\text{TLS-HS}}$ values for pure ML-KEM (0.92--0.98$\times$) fall below 1.0, we have shown that these differences are not statistically significant at the TLS protocol level. \par}

\subsection{Impact of Response Payload Size on PQC Overhead}
\label{sec:payload_impact}

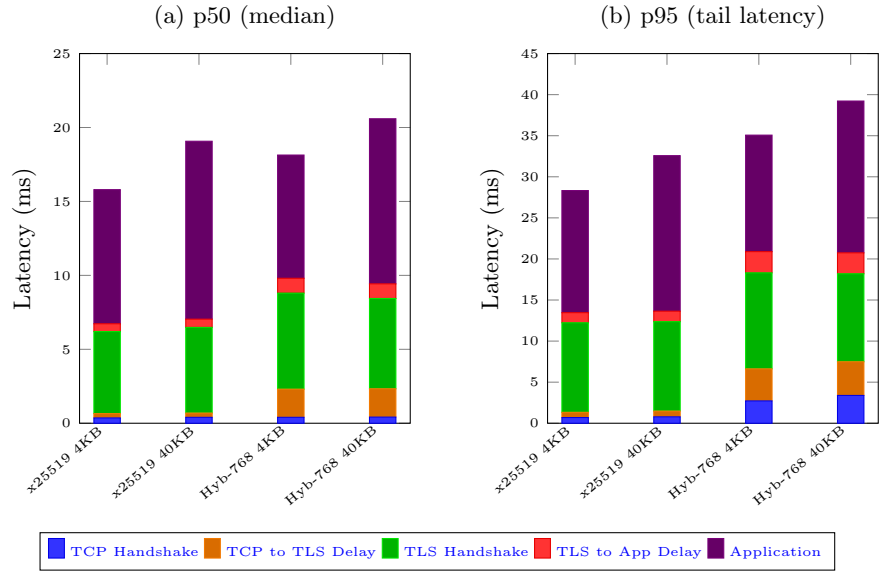
\begin{figure}[H]
\centering
\begin{minipage}[b]{0.49\textwidth}
\centering
\begin{tikzpicture}
\begin{axis}[
    ybar stacked, bar width=10pt,
    width=\textwidth, height=6.5cm,
    title={\footnotesize (a) p50 (median)},
    ylabel={\footnotesize Latency (ms)},
    symbolic x coords={x25519 4KB,x25519 40KB,Hyb-768 4KB,Hyb-768 40KB},
    xtick=data, x tick label style={rotate=40, anchor=east, font=\tiny},
    ymin=0, ymax=25,
    legend to name=payloadlegend,
    legend style={font=\tiny, legend columns=5},
    tick label style={font=\tiny},
    ytick={0,5,...,25},
]
\addplot+[fill=blue!80, draw=blue!90!black] coordinates {(x25519 4KB,0.360) (x25519 40KB,0.401) (Hyb-768 4KB,0.402) (Hyb-768 40KB,0.417)};
\addplot+[fill=orange!85!black, draw=orange!95!black] coordinates {(x25519 4KB,0.294) (x25519 40KB,0.289) (Hyb-768 4KB,1.903) (Hyb-768 40KB,1.915)};
\addplot+[fill=green!70!black, draw=green!90!black] coordinates {(x25519 4KB,5.547) (x25519 40KB,5.791) (Hyb-768 4KB,6.495) (Hyb-768 40KB,6.105)};
\addplot+[fill=red!80, draw=red!90!black] coordinates {(x25519 4KB,0.526) (x25519 40KB,0.559) (Hyb-768 4KB,1.004) (Hyb-768 40KB,0.989)};
\addplot+[fill=violet!80!black, draw=violet!95!black] coordinates {(x25519 4KB,9.071) (x25519 40KB,12.032) (Hyb-768 4KB,8.334) (Hyb-768 40KB,11.166)};
\legend{TCP Handshake, TCP to TLS Delay, TLS Handshake, TLS to App Delay, Application}
\end{axis}
\end{tikzpicture}
\end{minipage}
\hfill
\begin{minipage}[b]{0.49\textwidth}
\centering
\begin{tikzpicture}
\begin{axis}[
    ybar stacked, bar width=10pt,
    width=\textwidth, height=6.5cm,
    title={\footnotesize (b) p95 (tail latency)},
    ylabel={\footnotesize Latency (ms)},
    symbolic x coords={x25519 4KB,x25519 40KB,Hyb-768 4KB,Hyb-768 40KB},
    xtick=data, x tick label style={rotate=40, anchor=east, font=\tiny},
    ymin=0, ymax=45,
    tick label style={font=\tiny},
    ytick={0,5,10,...,45},
]
\addplot+[fill=blue!80, draw=blue!90!black] coordinates {(x25519 4KB,0.694) (x25519 40KB,0.781) (Hyb-768 4KB,2.720) (Hyb-768 40KB,3.386)};
\addplot+[fill=orange!85!black, draw=orange!95!black] coordinates {(x25519 4KB,0.635) (x25519 40KB,0.702) (Hyb-768 4KB,3.900) (Hyb-768 40KB,4.097)};
\addplot+[fill=green!70!black, draw=green!90!black] coordinates {(x25519 4KB,10.903) (x25519 40KB,10.886) (Hyb-768 4KB,11.692) (Hyb-768 40KB,10.715)};
\addplot+[fill=red!80, draw=red!90!black] coordinates {(x25519 4KB,1.227) (x25519 40KB,1.272) (Hyb-768 4KB,2.576) (Hyb-768 40KB,2.546)};
\addplot+[fill=violet!80!black, draw=violet!95!black] coordinates {(x25519 4KB,14.866) (x25519 40KB,18.935) (Hyb-768 4KB,14.186) (Hyb-768 40KB,18.479)};
\end{axis}
\end{tikzpicture}
\end{minipage}
\par\vspace{4pt}
\ref{payloadlegend}
\caption{End-to-end latency decomposition comparing 4\,KB and 40\,KB backend response bodies for x25519 (classical) and x25519\_MLKEM768 (hybrid). Left: p50. Right: p95.}
\label{fig:payload_comparison}
\end{figure}

To evaluate whether the cryptographic overhead introduced by post-quantum algorithms scales with application-layer payload size, additional tests were conducted using a 40\,KB backend response body for x25519 (classical) and x25519\_ MLKEM768 (hybrid). These results can be compared directly against their 4\,KB counterparts reported in previous sections. Table~\ref{tab:latency_summary} includes the full per-layer latency breakdown for both payload sizes, and Figure~\ref{fig:payload_comparison} provides a stacked visualization of the end-to-end latency decomposition across the four configurations.

\medskip
\textbf{TCP Handshake layer unaffected.} The TCP layer show no statistically meaningful variation between the 4\,KB and 40\,KB scenarios for either algorithm.

\medskip
\textbf{Application layer dominance.} Increasing the response body from 4\,KB to 40\,KB raises the median application response latency from 9.07\,ms to 12.03\,ms for x25519 (+32.7\%) and from 8.33\,ms to 11.17\,ms for x25519\_MLKEM768 (+34.0\%). This increase is attributable entirely to the larger payload transfer and end-to-end network path delay; no interaction with the cryptographic algorithm is observed.

\medskip
\textbf{Cryptographic layers unaffected.} The TCP-to-TLS delay, TLS handshake, and TLS-to-application layers show no statistically meaningful variation between the 4\,KB and 40\,KB scenarios for either algorithm. For x25519\_ MLKEM768, the median TCP-to-TLS delay is 1.903\,ms (4\,KB) versus 1.915\,ms (40\,KB); the TLS handshake latency is 6.495\,ms versus 6.105\,ms; and the TLS-to-application delay is 1.004\,ms versus 0.989\,ms. These differences are within measurement noise and confirm that the PQC overhead is entirely \emph{front-loaded} during connection establishment and independent of subsequent data transfer volume.

\medskip
{\sloppy\textbf{Dilution of relative PQC overhead.} From an end-to-end perspective, the absolute PQC overhead (the difference between x25519\_MLKEM768 and x25519) remains approximately constant: 2.34\,ms at the median for the 4\,KB scenario versus 1.52\,ms for the 40\,KB scenario. However, because the total connection time increases with payload size, the \emph{relative} overhead decreases from approximately 14.8\% (4\,KB) to 8.0\% (40\,KB) at the median. At p95, the absolute overhead is 6.75\,ms (4\,KB) versus 6.65\,ms (40\,KB), while the relative overhead drops from 23.8\% to 20.4\%. This dilution effect implies that in real-world applications with substantial response payloads, the performance penalty of PQC migration becomes proportionally less significant.\par}

\medskip
\textbf{Scope of payload analysis.} The 40\,KB evaluation was conducted only for x25519 and x25519\_MLKEM768 (hybrid). Extending this analysis to pure ML-KEM configurations (MLKEM512, MLKEM1024) with larger payload sizes is planned for future work.

The cryptographic layers (TCP-to-TLS, TLS handshake, TLS-to-App) remain essentially unchanged with increasing payload size, while the application layer grows proportionally. The relative PQC overhead diminishes as the payload-dependent portion of the total latency increases.
\subsection{CPU and Network Utilization}
\label{sec:cpu_net}

\begin{figure}[H]
\centering
\begin{minipage}[b]{0.54\textwidth}
\centering
\begin{tikzpicture}
\begin{axis}[
    ybar, bar width=9pt,
    width=\textwidth, height=5.5cm,
    ylabel={\footnotesize CPU utilization (\%)},
    symbolic x coords={x25519,Hyb-512,Hyb-768,ML-512,ML-1024},
    xtick=data, x tick label style={rotate=30, anchor=east, font=\tiny},
    legend style={at={(0.02,0.98)}, anchor=north west, font=\tiny},
    ymin=0, ymax=22,
    enlarge x limits=0.15,
    tick label style={font=\tiny},
]
\addplot+[fill=blue!50] coordinates {(x25519,3.72) (Hyb-512,8.24) (Hyb-768,8.71) (ML-512,7.91) (ML-1024,7.82)};
\addplot+[fill=red!50] coordinates {(x25519,14.61) (Hyb-512,15.38) (Hyb-768,15.46) (ML-512,13.91) (ML-1024,14.34)};
\legend{Client (Sys+User), Server (Sys+User)}
\end{axis}
\end{tikzpicture}
\end{minipage}
\caption{Total CPU utilization (Sys+User) per algorithm (backend 4\,KB)}
\label{fig:cpu_network}
\end{figure}
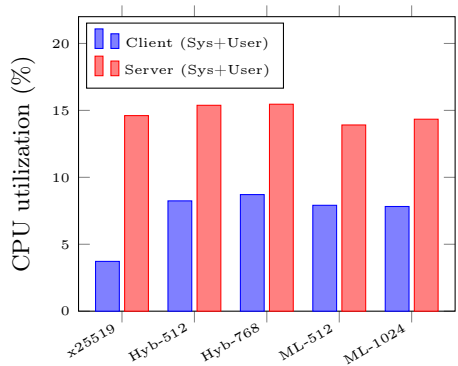

Table~\ref{tab:cpu_network} summarizes CPU utilization and network overhead in cryptographic configurations, while Figure~\ref{fig:cpu_network} depicts graphically the CPU utilization .

\begin{table}[H]
\centering
\caption{Mean CPU utilization and network overhead per cryptographic configuration (backend 4\,KB). Client and Server CPU show total utilization (System + User).}
\label{tab:cpu_network}
\resizebox{\textwidth}{!}{%
\begin{tabular}{lcccc}
\hline
\textbf{Algorithm} & \textbf{Client CPU (\%)} & \textbf{Server CPU (\%)} & \textbf{Traffic (Mb/s)} & \textbf{key\_share (B)} \\
\hline
x25519 & 3.72 & 14.61 & 5.18 & 32 \\
x25519\_MLKEM512 & 8.24 & 15.38 & 5.88 & 832 \\
x25519\_MLKEM768 & 8.71 & 15.46 & 6.14 & 1216 \\
MLKEM512 & 7.91 & 13.91 & 5.83 & 800 \\
MLKEM1024 & 7.82 & 14.34 & 6.55 & 1568 \\
\hline
\end{tabular}%
}
\end{table}

{\sloppy Client-side CPU utilization approximately doubles under hybrid and post-quantum configurations, reflecting the additional cost of TLS initialization and key exchange preparation. Server-side CPU utilization shows a modest absolute variation ($\sim$14--15\%), but this must be interpreted in relative terms: hybrid configurations increase server CPU from 14.61\% (x25519) to 15.46\% (x25519\_MLKEM768), representing a \textbf{+5.8\% relative increase}. Pure ML-KEM variants show slightly \emph{lower} server CPU (13.91\% for MLKEM512, $-4.8\%$ relative), although the absolute differences are small and may fall within system-level variability. Although the server operates well below saturation at 100~TPS---a deliberate choice to avoid multithreading contention artifacts in the virtualized test environment---\textbf{extrapolating these relative differences to higher load levels suggests that PQC server-side overhead could become significant under production-scale traffic}.\par}

{\sloppy\textbf{Why the server-side cost is small.} The modest server-side variation is expected rather than anomalous, and follows from the computational asymmetry of ML-KEM key exchange in TLS~1.3. In the ephemeral ML-KEM design used here (per the IETF hybrid key-exchange draft~\cite{ietf_hybrid} and FIPS~203~\cite{nist_mlkem}), the \emph{client} generates a fresh ephemeral KEM key pair and transmits its public key in the ClientHello \texttt{key\_share}, then decapsulates the shared secret on receipt of the ServerHello; the \emph{server} performs only a single encapsulation against the client-supplied public key. Key generation and decapsulation---the heavier client-side operations---are precisely what surfaces in the TCP-to-TLS delay, whereas encapsulation is comparatively inexpensive. Importantly, the server does \emph{not} generate a fresh, signed KEM key pair per handshake (as in KEMTLS-style proposals~\cite{schwabe2024}); server authentication is provided by the unchanged RSA-2048 signature across all configurations, so the only added server work is the lightweight encapsulation. This confirms that the server was configured correctly and that the low observed cost reflects the genuine efficiency of ML-KEM encapsulation rather than a misconfiguration.\par}

Network traffic scales linearly with key exchange size, from 32 bytes for x25519 up to 1568 bytes for MLKEM1024. Despite this increase, no proportional impact on latency is observed, indicating that the bandwidth overhead introduced by post-quantum key exchanges remains manageable in the evaluated architecture.

\medskip
\sloppy{\textbf{Implications for Different Client Types.}
The asymmetric CPU impact---where client-side utilization approximately doubles while the server shows a moderate but non-negligible relative increase of up to 5.8\% for hybrids PQC ---has important implications for PQC migration planning. On the client side, the overhead is substantial and may become a bottleneck for resource-constrained devices such as IoT sensors, mobile handsets, or embedded systems. On the server side, although absolute CPU differences are small ($\sim$1\%), the relative impact is meaningful: at the tested load of 100~TPS the server operates at only $\sim$15\% CPU, so a 1\% absolute increase represents $\sim$6--7\% of the utilized capacity. Under production workloads that drive servers closer to saturation, this relative overhead would translate into proportionally larger absolute increases. }

More importantly when evaluating network devices that are acting as MiTM (Man-in-The-Middle) performing traffic decryption, inspection and then re-encryption of the traffic, the expected impact should be on the highest side. For such network devices a collaborative effect of both client side and server side PQC impact in addition to high TLS session rates can create the conditions of signifiant performance degradation when migrating to PQC encryption.
Practitioners should therefore evaluate PQC overhead at client, server side, and more importantly at network level with representative traffic load levels, before deployment\textbf{.
}
\subsection{End-to-End Perspective}
\label{sec:e2e}

The preceding per-layer analysis (Sections~\ref{sec:results:lat_breakdown}--\ref{sec:cpu_net}) demonstrated that PQC overhead is concentrated in the TCP-to-TLS delay (ClientHello construction), while the TLS handshake exchange, application, and TCP layers remain effectively algorithm-neutral. We now consolidate these findings into aggregate end-to-end connection times (TCP~SYN to HTTP~200~OK), drawing on the stacked decomposition shown in Figure~\ref{fig:stacked_layers} and the per-layer data in Table~\ref{tab:latency_summary}.

Table~\ref{tab:e2e_summary} reports the end-to-end latency percentiles for each configuration, computed directly across all valid TCP streams in each capture (i.e., the true per-connection total time from TCP~SYN to HTTP~200~OK, not the sum of per-layer percentiles, which would be statistically invalid since percentiles are not additive).

\begin{table}[H]
\centering
\caption{End-to-end (E2E) latency (TCP~SYN to HTTP~200~OK): percentiles and standard deviation per algorithmic configuration and backend response size. Values are computed directly from per-connection total times, not from summing per-layer percentiles.}
\label{tab:e2e_summary}
\begin{tabular}{lcccc}
\hline
\textbf{Configuration} & \textbf{p50 (ms)} & \textbf{p95 (ms)} & \textbf{p99 (ms)} & \textbf{SD (ms)} \\
\hline
x25519 (4\,KB) & 16.54 & 23.41 & 26.15 & 4.93 \\
x25519 (40\,KB) & 19.88 & 27.10 & 30.37 & 30.87 \\
x25519\_MLKEM512 (4\,KB) & 20.26 & 26.91 & 29.70 & 5.70 \\
x25519\_MLKEM768 (4\,KB) & 19.63 & 26.14 & 28.30 & 6.42 \\
x25519\_MLKEM768 (40\,KB) & 22.25 & 29.89 & 33.40 & 21.17 \\
MLKEM512 (4\,KB) & 18.92 & 25.53 & 28.08 & 6.15 \\
MLKEM1024 (4\,KB) & 19.16 & 25.58 & 27.73 & 5.40 \\
\hline
\end{tabular}
\end{table}

\textbf{Key observations.} Under the 4\,KB backend scenario, the classical baseline (x25519) achieves a median end-to-end latency of 16.54\,ms (SD\,=\,4.93\,ms). Hybrid configurations increase this to 20.26\,ms (x25519\_MLKEM512, $+$22.5\%) and 19.63\,ms (x25519\_MLKEM768, $+$18.7\%), while pure ML-KEM variants reach 18.92\,ms (MLKEM512, $+$14.4\%) and 19.16\,ms (MLKEM1024, $+$15.8\%). Notably, MLKEM1024 achieves virtually the same end-to-end latency as MLKEM512 despite its higher security level, consistent with the per-layer finding that the TLS handshake exchange is algorithm-neutral.

The absolute PQC overhead at the median ranges from 2.38\,ms (pure ML-KEM) to 3.72\,ms (hybrid) relative to x25519. As shown in Figure~\ref{fig:stacked_layers}, this overhead is almost entirely attributable to the TCP-to-TLS layer ($\sim$1.5\,ms increase) with minor contributions from the TLS-to-App delay ($\sim$0.4\,ms). The TLS handshake and application layers contribute negligibly to the inter-algorithm differences. At the tail (p95), the overhead narrows: 2.12--3.50\,ms across PQC configurations, indicating that the penalty is relatively stable and does not amplify disproportionately under tail conditions.

For the 40\,KB scenario, the standard deviations increase substantially (30.87\,ms for x25519, 21.17\,ms for x25519\_MLKEM768), reflecting occasional long transfers at the application layer. Nevertheless, the absolute PQC overhead at the median remains stable at approximately 2.37\,ms, and the relative overhead decreases from $\sim$19\% (4\,KB) to $\sim$12\% (40\,KB) due to the dilution effect described in Section~\ref{sec:payload_impact}.

\section{Conclusions and Future Work}
This paper presented a layered performance analysis of TLS~1.3 handshakes comparing classical (x25519), hybrid (x25519+ML-KEM), and pure post-quantum (ML-KEM) key exchange configurations under realistic load conditions of 100 requests per second. Through more than thirty experiments and statistical analysis of each protocol layer, we draw the following conclusions.

\textbf{TCP layer independence (Section~\ref{sec:results:TCP-Handshake}).} The TCP handshake latency remains unaffected by the choice of cryptographic algorithm: median values range from 0.360 to 0.402\,ms, a maximum difference of 0.042\,ms ($\Delta=0.17$, negligible). This confirms that PQC overhead does not propagate to lower protocol layers.

{\sloppy\textbf{Client-side initialization as the dominant PQC cost (Section~\ref{sec:results:TCP-to-TLS}).} The TCP-to-TLS delay (SYN-ACK to ClientHello) represents the most significant performance penalty introduced by post-quantum algorithms. Median latency increases from 0.294\,ms (x25519, SD\,=\,0.194\,ms) to 1.73--1.90\,ms for all PQC variants---a $6\times$ increase ($\Delta\approx7.7$, far exceeding Cohen's large-effect threshold of~0.8), making this the only layer where the PQC overhead is unambiguously significant. This cost is driven primarily by client-side key material generation and ClientHello construction, independent of backend response size and consistent across ML-KEM security levels.\par}

{\sloppy\textbf{TLS handshake exchange as an algorithm-neutral phase (Section~\ref{sec:results:TLS-Handshake}).} The TLS handshake exchange latency (ClientHello to Finished) shows no practically significant variation across algorithms: all configurations fall within 5.253--6.495\,ms at the median, with differences of 0.1--0.9\,ms yielding $\Delta=0.03$--$0.33$ (negligible to small per Cohen's thresholds~\cite{cohen1988}). We stress that this neutrality concerns the on-the-wire message exchange, \emph{not} the ClientHello construction, whose algorithm-sensitive cost is captured separately in the TCP-to-TLS layer. While pure ML-KEM microbenchmarks demonstrate faster encapsulation/decapsulation than ECDH scalar multiplication~\cite{bos2018,paquin2020}, this advantage is indistinguishable from noise in our TLS measurements. The practical implication is that pure PQC key exchange introduces \emph{no measurable penalty in the handshake exchange} compared to the classical baseline at this load level (see Section~\ref{sec:normalized_overhead} for normalized metrics).\par}

\textbf{Negligible application-layer impact (Section~\ref{sec:results:TLS-to-App_App}).} Application response latency is entirely dominated by the end-to-end network path latency and response payload size (approximately 9\,ms for 4\,KB and 12\,ms for 40\,KB responses). The maximum inter-algorithm variation (0.74\,ms) yields $\Delta=0.21$ (negligible to small), confirming no practically meaningful penalty attributable to the cryptographic algorithm. PQC migration does not degrade user-perceived application performance.

\textbf{Moderate normalized cryptographic overhead (Section~\ref{sec:normalized_overhead}).} Despite the $\sim$$6\times$ overhead in the TCP-to-TLS layer, the Cryptographic Overhead Share (COS) remains moderate: 6--14\% of the total end-to-end connection time across all PQC configurations. Hybrid algorithms exhibit the highest COS (10.6--14.1\%), while pure ML-KEM configurations achieve 6.0--7.9\%. This confirms that the absolute PQC penalty, although concentrated in a single layer, has a limited impact on overall transaction latency.

\textbf{Payload size dilution (Section~\ref{sec:payload_impact}).} Increasing the backend response body from 4\,KB to 40\,KB does not affect the cryptographic layers (TCP-to-TLS, TLS handshake, TLS-to-App remain unchanged), while the application layer grows proportionally. Consequently, the relative PQC overhead decreases from $\sim$15\% (4\,KB) to $\sim$8\% (40\,KB) at the median, as the payload-dependent latency dilutes the fixed cryptographic cost.

\textbf{CPU and network overhead (Section~\ref{sec:cpu_net}).} Client-side CPU approximately doubles under PQC (from ${\sim}3.7\%$ to ${\sim}8.2$--$8.7\%$), while server-side CPU remains within a narrow range (${\sim}14$--$15\%$), with hybrid configurations showing up to $\sim$6\% relative increase. Network traffic grows modestly from 5.2 to 6.6\,Mb/s despite key exchange sizes increasing from 32\,B (x25519) to 1568\,B (MLKEM1024). PQC migration planning should consider client-side resource provisioning (especially for constrained devices), network devices, and server-side capacity at production load levels.

\textbf{End-to-end perspective (Section~\ref{sec:e2e}).} The PQC penalty on the end-to-end (E2E) transaction time is a \emph{fixed} cost concentrated in connection establishment: PQC adds approximately 2--4\,ms regardless of payload size, because the post-quantum cost is incurred once during ClientHello construction and is independent of the volume of application data subsequently transferred. The E2E time therefore depends on both the key-exchange algorithm \emph{and} the payload size, but the post-quantum component does not grow with message size. Consequently, the \emph{relative} weight of this fixed overhead shrinks as the payload (and thus the total transaction time) grows: in direct scenarios it amounts to roughly 15--20\% of the E2E time, whereas in backend-proxied scenarios with larger cleartext responses it falls below 20\% and continues to decrease with payload size.

\medskip
Given the findings from this paper, there are few important feature research directions that are important to analyse in details including:
\begin{itemize}
  \item extending the analysis to real network environments with commercial load balancers and MiTM (Man-in-The-Middle) inspection devices to quantify the performance impact when using PQC in TLS especially considering the high network load/connections these devices need to handle, where PQC performance impact may be signifiant.
  \item conducting stress tests beyond 100 TPS to identify the breaking point of PQC-enabled TLS under heavy load.
  \item evaluating post-quantum digital signature algorithms (Falcon, SPHINCS+, ML-DSA) and their impact on certificate verification latency.
  \item extending the comparison to further key-exchange algorithms and parameter sets beyond the five evaluated here---for example, additional ML-KEM security levels (e.g., MLKEM768 in pure mode), the SecP256r1MLKEM768 and SecP384r1MLKEM1024 hybrid groups, and alternative KEM families---to broaden the coverage of the post-quantum key-exchange design space.
  \item fine-grained profiling of the individual cryptographic operations within the TCP-to-TLS layer (ephemeral key-pair generation, encapsulation, and decapsulation) to attribute the client-side overhead to specific primitives and identify concrete targets for code-level optimization.
  \item isolating the resource cost (CPU and bandwidth) attributable \emph{exclusively} to the TLS handshake, decoupled from subsequent application-record delivery, to characterize the handshake-only footprint of hybrid versus pure PQC configurations.
  \item evaluating PQC TLS overhead in stacks with native PQC support, such as Nginx with OpenSSL~3.5 and Locust with OpenSSL~3.5.
\end{itemize}

\section*{Acknowledgments}
In accordance with the conference policy on AI-generated content, we disclose the following. An AI coding assistant (GitHub Copilot, powered by large language models) was used partially during two activities in this work: (1)~\emph{Script development}: the data-reduction tool (\texttt{pcap\_layer\_analysis.py}) was co-developed with AI assistance for code generation, refactoring, and debugging; and (2)~\emph{Paper drafting}: AI was used as a writing aid in the preparation of selected sections of this manuscript, including text generation, restructuring, and grammar refinement. In both cases, all AI-generated output was reviewed, validated, and edited by the authors, who bear full responsibility for the final content. 

This work was supported by the PQNEXT project, which has received funding from the European Union’s Horizon Europe research and innovation programme under Grant Agreement No. 101225759.

%
%
%
%

\end{document}